# Negative Compressibility in Graphene-terminated Black Phosphorus Heterostructures

Yingying Wu[1,†], Xiaolong Chen[1,2,†], Zefei Wu[1], Shuigang Xu[1], Tianyi Han[1], Jiangxiazi Lin[1], Yuan Cai[1], Yuheng He[1], Chun Cheng[3], Ning Wang*,[1]

[1]*Department of Physics and the William Mong Institute of Nano Science and Technology, the Hong Kong University of Science and Technology, Hong Kong, China*

[2] *Department of Engineering and Cambridge Graphene Centre, University of Cambridge, 9, JJ Thomson Avenue, Cambridge, United Kingdom, CB3 0FA*

[3] *Department of Materials Science and Engineering and Shenzhen Key Laboratory of Nanoimprint Technology, South University of Science and Technology, Shenzhen 518055, China*

**Keywords:** Black phosphorus, negative compressibility, many-body effect, quantum capacitance, encapsulation, heterostructure.

**Negative compressibility generated by many-body effects in 2D electronic systems can enhance gate capacitance. We observe capacitance enhancement in a newly emerged 2D layered material, atomically thin black phosphorus (BP). The encapsulation of BP by hexagonal boron nitride sheets with few-layer graphene as a terminal ensures ultraclean heterostructure interfaces, allowing us to observe negative compressibility at low hole carrier concentrations. We explained the negative compressibility based on the Coulomb correlation among in-plane charges and their image charges in a gate electrode in the framework of Debye screening.**

Negative compressibility in an electronic system describes the effect of electron-electron (e-e) interaction, which lowers the potential of electrons when the carrier density of a system increases[1-4]. This effect provides a new strategy to enhance the gate capacitance of field-effect transistors (FETs) beyond the expected geometric capacitance for low-power dissipation[4-8]. The effects of negative compressibility have been previously observed in ultraclean systems, such as in high-quality LaAlO$_3$/SrTiO$_3$ interfaces[9], two-dimensional (2D) GaAs systems[1,10-12] and ferroelectric materials[13], where Coulomb interactions are normally strong and play an important role in transport properties. Newly emerged 2D layered semiconductors, such as transition metal dichalcogenides[14-18] and black phosphorus (BP)[19-28], are new platforms for both nanotechnology and fundamental physics. Among these 2D materials, atomically thin BP is a promising channel material of FETs with high mobility[19,23,28,29] and high stability by encapsulating BP with hexagonal boron nitride (BN) sheets. Our previous work has found that high-quality BN/BP/BN heterostructures with clean interfaces are ideal 2D electron systems with high mobility[19]. High-quality BN/BP/BN heterostructures also offer great opportunities to investigate quantum Hall effects[23], as well as Coulomb interactions of charge carriers in 2D materials.

In this work, we report the observation of negative electronic compressibility in BN/BP/BN capacitors constructed using few-layer graphene as terminal electrodes. Gate capacitance enhancement is apparent near the depletion region. The observation of this enhancement effect largely relies on ultraclean interfaces in BN-BP-BN heterostructures and very low impurity concentrations.

A simple parallel-plate capacitor structure is adopted to investigate the properties of BP. When both electrodes in a standard parallel plate capacitor are made of normal metals, the total capacitance $C_t$ purely originates from the geometric capacitance $C_g = \frac{\varepsilon}{d}$ because the metal electrodes can perfectly screen the electric field, where $\varepsilon$ and $d$ are the dielectric constant and thickness of the capacitor, respectively. However, if one of the electrodes is replaced with a 2D material with finite Debye screening radius $R_D$, the electric field penetrates distance $R_D$ into the electrode[30,31], thereby decreasing total capacitance $C_t = \frac{\varepsilon}{d + R_D}$. Capacitance would be enhanced if the screening radius is $R_D < 0$. At a low carrier density, a negative Debye screening length can occur when electron-electron interactions are sufficiently strong, whereas the occurrence of disorder truncates the enhancement of the capacitance [9,10,32].

The polymer-free dry transfer method[33] is adopted to assemble exfoliated few-layer BP, BN, and graphene flakes into a BN-BP-BN heterostructure with few-layer graphene as a terminal electrode, as shown in Fig. 1a and 1c. Annealing treatments were then conducted on the BN-BP-BN heterostructure at high temperatures up to 500 °C before performing metal electrode deposition. BN encapsulation and annealing treatment guarantee a clean BN-BP interface with low impurity concentrations[19]. At last, Cr/Au (2 nm/60 nm) films were deposited to form a top gate and an ohmic contact with graphene. Polarized Raman spectroscopy was performed to characterize the properties of the few-layer BP samples. The angle-dependent Raman spectra obtained from the BP samples are shown in Fig. 1d, which agree with previous studies[34-36].

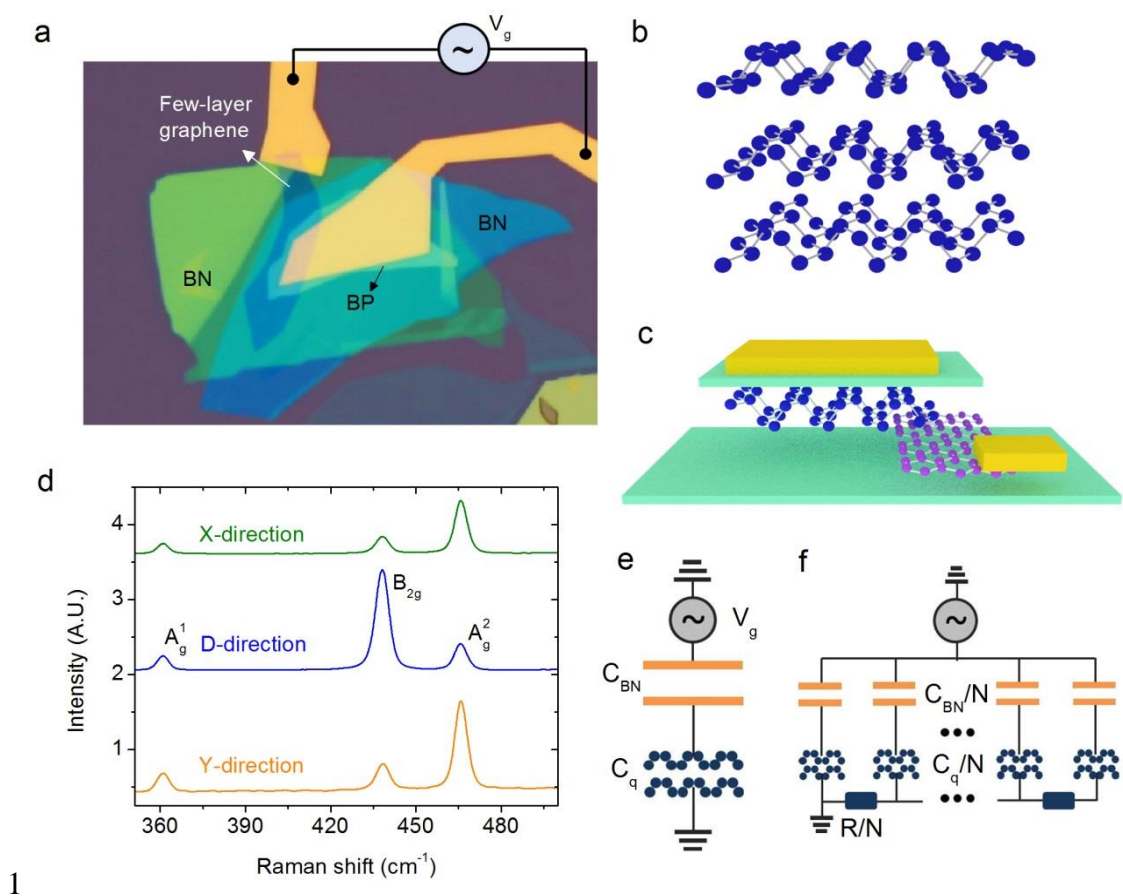

FIG. 1 BP vertical capacitor heterostructure. (a) False-color optical and (c) schematic image of the BN-BP-BN heterostructure with few-layer graphene as terminal electrode. The BP flake is 8 nm thick, as determined using an atomic force microscope. (b) A schematic image of few-layer BP. (d) Determining the crystal orientation via Raman spectroscopy. The intensity values were normalized to the Si peak from the substrate. (e, f) Equivalent circuits of the measurement setup (e) with and (f) without taking into account the resistance of BP.

Using a graphene terminal to contact the BN-BP-BN heterostructure has several benefits compared with other structures (see Supplementary Fig. 3). This graphene terminal does not only simplify the fabrication procedures without a tedious chemical etching process[37] to make contact with the encapsulated BP flakes but also ensures a polymer-free BN-BP clean interface.

The capacitance is measured between the top gate and the graphene terminal with an AC excitation frequency ranging from 20 Hz to 1 MHz (see Fig. 1a). By applying a top gate voltage $V_g$, the Fermi energy of BP can be tuned continuously. The measured capacitance $C_t$ is the total capacitance contributed by the two capacitors in a serial connection as shown in Fig. 1e. One of the capacitors is the constant geometric capacitance of the parallel-plate capacitor $C_g = \frac{\varepsilon_{BN}}{d}$, where $\varepsilon_{BN}$ is the dielectric constant of BN. The other one originates from the quantum capacitance of BP $C_q = e^2 \frac{dn}{d\mu}$ [35,38,39], where $n$ and $\mu$ are the carrier density and chemical potential, respectively. Here, $\frac{dn}{d\mu}$ is the electronic compressibility, which describes the ability of an electronic system to screen external charges [1,10,11,32,40]. For normal metals, the compressibility approaches infinity which results in a complete screening of external charges. By contrast, $\frac{dn}{d\mu}$ is a finite value in low-dimensional materials, partially penetrated by external electric field. Hence, total capacitance $C_t$ can be expressed as:

$$\frac{1}{C_t} = \frac{1}{C_g} + \frac{d\mu/dn}{e^2} \tag{1}$$

The measured capacitance of the BN-BP-BN heterostructure with an excitation frequency of 1 kHz is shown in Fig. 2a. Inside the bandgap of BP (gray region), the total capacitance approaches zero because of the vanishing density of the states. However, when $V_g$ is sufficiently large (blue region), the total capacitance approaches the geometric capacitance $C_t \approx C_g$, given that the large density of states would lead to a quantum capacitance $C_q$ much larger than $C_g$ [9]. The total capacitance can be expressed as $C_t = \frac{\varepsilon}{d^*} = \frac{\varepsilon}{d + R_D}$ by introducing an effective thickness of the system $d^*$.

Total capacitance $C_t$ is evidently enhanced near the depletion region of BP at low excitation frequencies ($f < 100\,\text{Hz}$). As shown in Fig. 2b, by decreasing the excitation frequencies from 200 Hz to 40 Hz, $C_t$ of the 5 nm-thick BP (Device A) gradually exceeds geometric capacitance $C_g$ and shows no sign of enhancement saturation within the measured frequency range. This capacitance enhancement at low excitation frequencies is also reproducible in other devices with different BP thicknesses, as shown in Fig. 2c (Device B with an 8 nm thick BP) and Fig. 2d (Device C with a 13 nm thick BP). Since the mobile carrier density is strongly suppressed near the depletion, $R$, the resistance of BP, can reach orders of $10^3\,\text{M}\Omega$ [37]. Therefore, the capacitor is not fully charged when the excitation frequency is $f > \frac{1}{RC_t}$; and thus $C_t$ is underestimated. The real capacitance can only be approached by lowering the excitation frequency. An equivalent circuit (Fig. 1f) can then be simplified into the two-serial-capacitor model shown in Fig. 1e when $f \ll \frac{1}{RC_t}$.

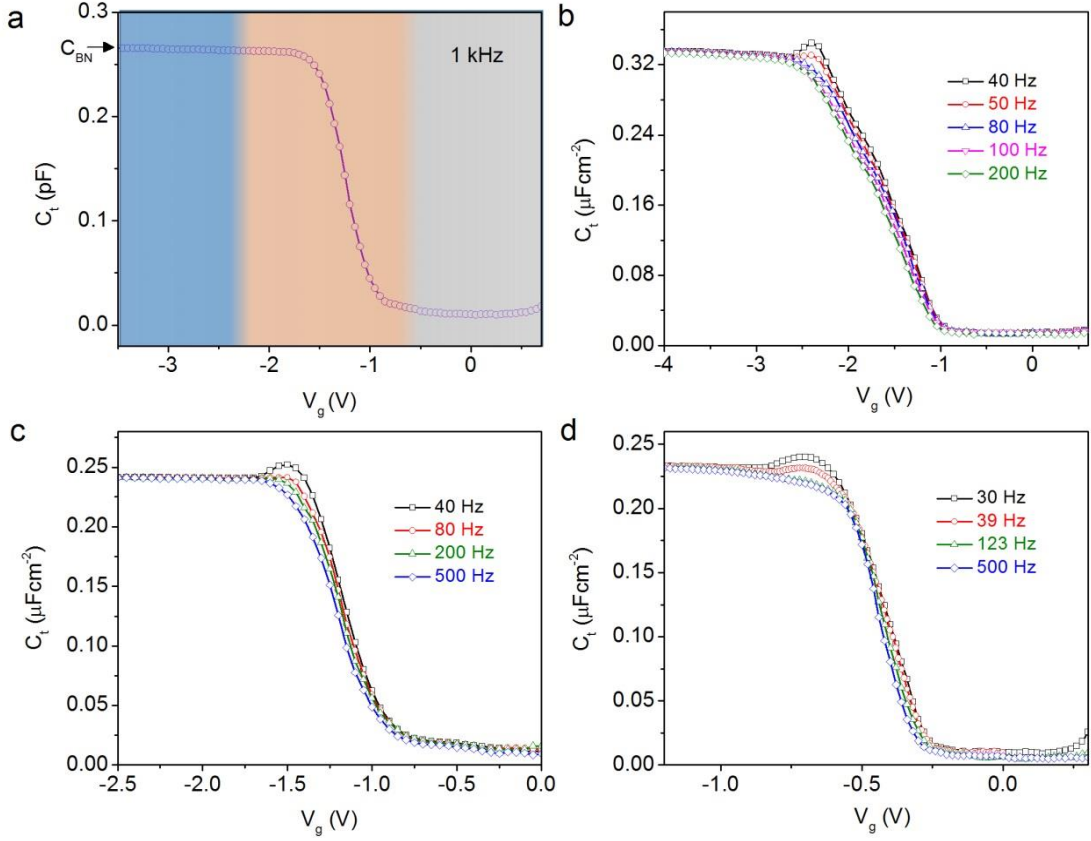

FIG. 2 Valence band of few-layer BP accessed by capacitance measurement. (a) Measured total capacitance of the 8 nm-thick sample in Fig.1a (device B) with an excitation frequency of 1 kHz at 2 K. (b-d) $C_t$ as a function of the gate voltage of (b) device A with 5 nm BP at 3K, (c) device B with 8 nm BP at 2K, and (d) device C with 13 nm BP at 2K.

The enhancement of $C_t$ ($C_t > C_g$) reveals that $C_q$ is negative, directly proving the negative compressibility. The correlation between inverse compressibility $\frac{d\mu}{dn}$ and carrier density $n = C_g(V_g - V_s - V_{th})/e$ can be obtained through Eq. (1), as shown in Fig. 3a, 3c and 3e, where $V_s$ and $V_{th}$ are the surface potential of BP and the threshold voltage, respectively. $V_s$ is extracted based on the charge conservation relation $V_s = \int_0^{V_g} (1 - \frac{C_t}{C_g}) dV_g$. For device A, negative compressibility occurs at a hole carrier density of ~$1.5 \times 10^{12}$ cm$^{-2}$, whereas negative compressibility occurs in devices B and C at a lower carrier density of below $0.8 \times 10^{12}$ cm$^{-2}$. This difference is caused by the quality variation in the samples. Given that thinner BP samples are more vulnerable to impurities, such samples require larger carrier densities to generate sufficient mobile carriers[41]. The Fermi energy $E_F = eV_s$ plotted as a function of gate voltage $V_g$ of three devices is shown in Fig. 3b, 3d and 3f. The Fermi energy generally increases with the gate voltage when $C_t < C_g$. Instead, an upturn is evident in the $E_F \sim V_g$ curve, which is a signature of negative compressibility[4,11].

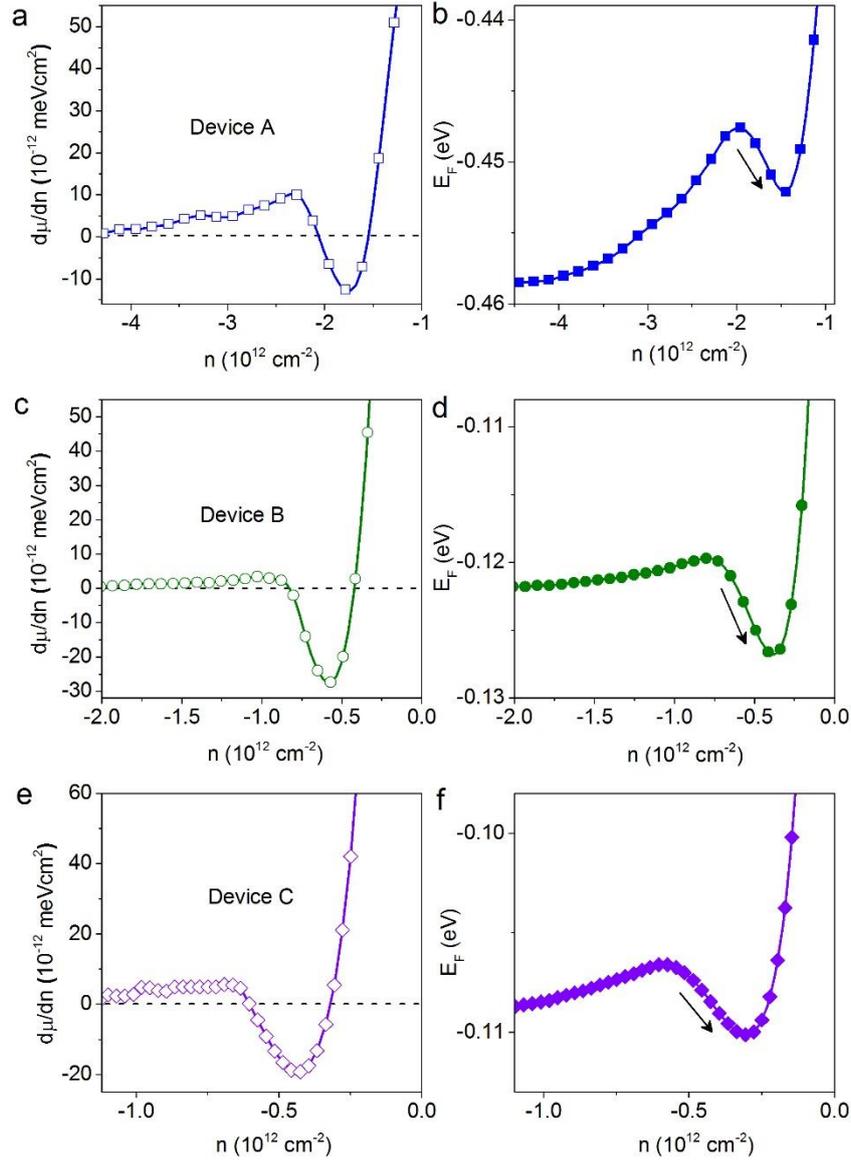

FIG. 3 Inverse of compressibility and the Fermi energy. (a,c,e) Inverse of compressibility of devices A, B, and C, respectively. The carrier density is determined by integrating the $C_t$-$V_g$ curve at the lowest excitation frequency, as shown in Fig. 2. The negative sign in the carrier density denotes the hole carrier density. (b,d,f) The $E_F \sim n$ relation at (b) 3K with an excitation frequency of 40Hz for device A, (d) 2K with an excitation frequency of 40Hz for device B, and (f) 2K with an excitation frequency of 30Hz for device C.

The negative compressibility in our system indicates the Debye screening length $R_D < 0$, which enhances the total capacitance $C_t = \dfrac{\varepsilon}{d + R_D}$. In the limit of the low-carrier density regime, where the average distance between two neighboring in-plane holes $n^{-1/2}$ is larger than the effective Bohr radius $a_0 = \varepsilon \hbar^2 / me^2$ of the holes in BP, 2D holes can form a strongly correlated system dominated by the Coulomb interaction[42]. Using BP dielectric constant $\varepsilon = 12\varepsilon_0$ [43] and effective mass of hole carriers $m \approx 0.35 m_e$ [22], a Bohr radius for BP of ~1.8 nm and the corresponding carrier density of $3.0 \times 10^{13}$ cm$^{-2}$ can be obtained. Hence, the BP system here (with a carrier density that is much smaller than $3.0 \times 10^{13}$ cm$^{-2}$) can be regarded as a strongly correlated system. In addition, the Coulomb interactions between the holes and the image charges of the holes in the gate electrode can also significantly enhance negative compressibility, especially when $n^{-1/2} \gg d$, where $n^{-1/2}$ is the average distance between two neighboring in-plane holes (see Fig. 4a)[32].

The theoretical corrections[32,44] to the effective thickness when considering the Coulomb interaction is as follows:

$$d^* \approx d - 0.12 n^{-1/2},\ (n^{1/2} d \gg 1) \qquad (2)$$

In the $n^{1/2} d \ll 1$ limit, effective thickness $d^*$ becomes very small, whereas the gate-capacitance enhancement could be huge, as shown in the following:

$$d^* = 2.7 d (nd^2)^{1/2},\ (n^{1/2} d \ll 1) \qquad (3)$$

Fig. 4b shows the experimental data of BP compared with theoretical function $d^*(d, n)$. The trend of the experimental data well agrees with the theoretical prediction. Notably, deviations occur, because disorder effects are not considered in the theoretical expressions, which screen Coulomb interactions among the holes. Negative compressibility could be suppressed when disorders prevail over electronic interactions. Thus only the negative compressibility of few-layer BP is observed in a region where $n^{-1/2}$ is relatively large and Coulomb interaction dominates.

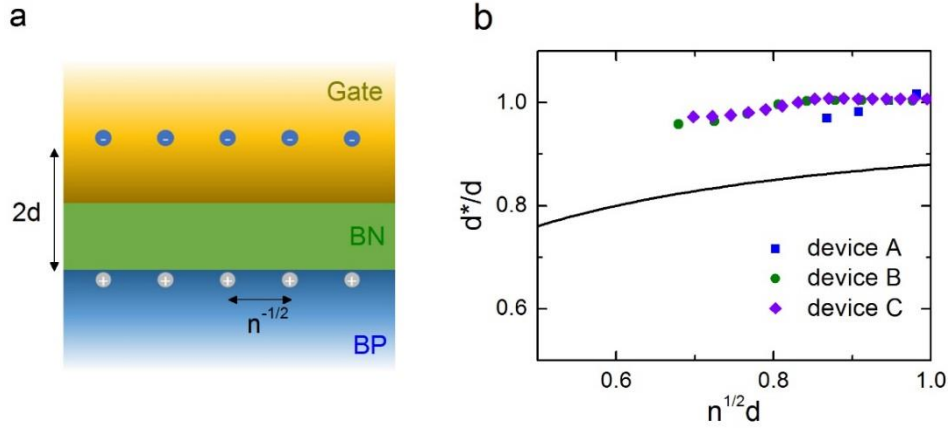

FIG. 4 Physical origin of negative compressibility detected in BP (a) Schematic coupling of the holes in BP to the image charges of the holes in the metal electrode. Neighboring holes (white circles) in the BP formed at the BP-BN interface. The distance between two neighboring holes is $n^{-1/2}$. Negative image charges (blue circles) are formed in the metal at a distance $2d$ away from the holes, where $d$ is the BN thickness. (b) The relation between ratio $\frac{d^*}{d}$ and dimensionless parameter $n^{1/2}d$. The black solid curve shows the theoretical prediction. The square, circle, and rhombus symbols represent the experimental data.

Aside from the Coulomb interaction among the carriers, other factors may also contribute to the observed negative compressibility. One possible factor is the frequency-dependent dielectric constant of the BN flakes. Larger geometric capacitance $C_g$ is generated when the dielectric constant of BN thin layers increases with decreased excitation frequency; hence, $C_t$ is enhanced. However, as shown in Fig. 2b-2d, the measured total capacitances at different low excitation frequencies for all three samples converge when the gate voltage is sufficiently large, which yields the same saturated values of $C_g$ at different excitation frequencies. Hence, capacitance enhancement is not caused by the frequency-dependent dielectric constant of BN layers. Another possible capacitance enhancement factor is excitation of localized electronic states in BP at low excitation frequencies. These localized states, which originate from defects or impurities, require a longer time (a smaller excitation frequency) to become excited. At low excitation frequencies, the excited localized states near the band edge could exhibit peak-like features in capacitance spectroscopy[45], which may be mistaken as the negative compressibility phenomena. However, the most distinct difference between them is that the peak-like capacitance features induced by localized states never exceed geometric capacitance $C_g$, disagreeing with experimental observations.

In conclusion, capacitance enhancement is discovered in high-quality BN-BP-BN heterostructures with few-layer graphene as terminal electrodes at sufficiently low temperatures and relatively low excitation frequencies. Negative electronic compressibility is evident near the band edge of few-layer BP. The negative compressibility observed proves the strong Coulomb interactions in few-layer BP in the low carrier density regime limit. Capacitance enhancement enables switching transistors at small gate voltages. High-quality BN-BP-BN heterostructures with negative compressibility also suggest potential applications of atomically thin materials for low-power nanoelectronics and optoelectronics.

**Author information**
Yingying Wu and Xiaolong Chen contribute equally to this work.

**Corresponding Author**
*E-mail: phwang@ust.hk

**Acknowledgment**
The authors are grateful for fruitful discussions about capacitance measurement technique with Professor Ho Bun Chan from the Physics Department, the Hong Kong University of Science and Technology. Financial support from the Research Grants Council of Hong Kong (Project Nos. 16302215, HKU9/CRF/13G, 604112, and N_HKUST613/12), the Guangdong Natural Science Funds for distinguished Young Scholar (Grand No. 2015A030306044)and technical support of the Raith-HKUST Nanotechnology Laboratory for the electron-beam lithography facility at MCPF (Project No. SEG_HKUST08) are hereby acknowledged.

**References**
[1]  G. Allison, E. A. Galaktionov, A. K. Savchenko, S. S. Safonov, M. M. Fogler, M. Y. Simmons, and D. A. Ritchie, Phys. Rev. Lett. **96**, 216407 (2006).
[2]  R. C. Ashoori and R. H. Silsbee, Solid State Commun. **81**, 821 (1992).
[3]  S. C. Dultz and H. W. Jiang, Phys. Rev. Lett. **84**, 4689 (2000).
[4]  S. V. Kravchenko, D. A. Rinberg, S. G. Semenchinsky, and V. M. Pudalov, Phys. Rev. B **42**, 3741 (1990).

[5] X. L. Chen *et al.*, Appl. Phys. Lett. **102**, 203103 (2013).

[6] S. Ilani, L. A. K. Donev, M. Kindermann, and P. L. McEuen, Nat Phys **2**, 687 (2006).

[7] A. K. Jonscher and M. N. Robinson, Solid-State Electronics **31**, 1277 (1988).

[8] S. Larentis, J. R. Tolsma, B. Fallahazad, D. C. Dillen, K. Kim, A. H. MacDonald, and E. Tutuc, Nano Lett. **14**, 2039 (2014).

[9] L. Li, C. Richter, S. Paetel, T. Kopp, J. Mannhart, and R. C. Ashoori, Science **332**, 825 (2011).

[10] J. P. Eisenstein, L. N. Pfeiffer, and K. W. West, Phys. Rev. Lett. **68**, 674 (1992).

[11] J. P. Eisenstein, L. N. Pfeiffer, and K. W. West, Phys. Rev. B **50**, 1760 (1994).

[12] A. L. Efros, F. G. Pikus, and V. G. Burnett, Phys. Rev. B **47**, 2233 (1993).

[13] A. I. Khan, K. Chatterjee, B. Wang, S. Drapcho, L. You, C. Serrao, S. R. Bakaul, R. Ramesh, and S. Salahuddin, Nat Mater **14**, 182 (2015).

[14] Q. H. Wang, K. Kalantar-Zadeh, A. Kis, J. N. Coleman, and M. S. Strano, Nat Nano **7**, 699 (2012).

[15] X. Xu, W. Yao, D. Xiao, and T. F. Heinz, Nat Phys **10**, 343 (2014).

[16] V. Podzorov, M. E. Gershenson, C. Kloc, R. Zeis, and E. Bucher, Appl. Phys. Lett. **84**, 3301 (2004).

[17] X. Cui *et al.*, Nat Nano **10**, 534 (2015).

[18] RadisavljevicB, RadenovicA, BrivioJ, GiacomettiV, and KisA, Nat Nano **6**, 147 (2011).

[19] X. Chen *et al.*, Nat Commun **6** (2015).

[20] H. Liu, A. T. Neal, Z. Zhu, Z. Luo, X. Xu, D. Tománek, and P. D. Ye, ACS Nano **8**, 4033 (2014).

[21] L. Li, Y. Yu, G. J. Ye, Q. Ge, X. Ou, H. Wu, D. Feng, X. H. Chen, and Y. Zhang, Nat Nano **9**, 372 (2014).

[22] J. Qiao, X. Kong, Z.-X. Hu, F. Yang, and W. Ji, Nat Commun **5** (2014).

[23] L. Li *et al.*, Nat Nano **10**, 608 (2015).

[24] R. Fei and L. Yang, Nano Lett. **14**, 2884 (2014).

[25] H. Yuan *et al.*, Nat Nano **advance online publication** (2015).

[26] S. P. Koenig, R. A. Doganov, H. Schmidt, A. H. Castro Neto, and B. Özyilmaz, Appl. Phys. Lett. **104**, 103106 (2014).

[27] G. Nathaniel *et al.*, 2D Materials **2**, 011001 (2015).

[28] Y. Cao *et al.*, arXiv preprint arXiv:1502.03755 (2015).

[29] N. Gillgren *et al.*, 2D Materials **2**, 011001 (2015).

[30] E. Stern, R. Wagner, F. J. Sigworth, R. Breaker, T. M. Fahmy, and M. A. Reed, Nano Lett. **7**, 3405 (2007).

[31] S. Das Sarma and B. A. Mason, Phys. Rev. B **31**, 5536 (1985).

[32] B. Skinner and B. I. Shklovskii, Phys. Rev. B **82**, 155111 (2010).

[33] L. Wang *et al.*, Science **342**, 614 (2013).

[34] S. Sugai and I. Shirotani, Solid State Commun. **53**, 753 (1985).

[35] F. Xia, H. Wang, and Y. Jia, Nat Commun **5** (2014).

[36] S. Zhang *et al.*, ACS Nano **8**, 9590 (2014).

[37] A. Avsar, I. J. Vera-Marun, J. Y. Tan, K. Watanabe, T. Taniguchi, A. H. Castro Neto, and B. Özyilmaz, ACS Nano **9**, 4138 (2015).

[38] S. Luryi, Appl. Phys. Lett. **52**, 501 (1988).

[39] J. Xia, F. Chen, J. Li, and N. Tao, Nat Nano **4**, 505 (2009).

[40] S. Ilani, A. Yacoby, D. Mahalu, and H. Shtrikman, Science **292**, 1354 (2001).

[41] T. Low, R. Roldán, H. Wang, F. Xia, P. Avouris, L. M. Moreno, and F. Guinea, Phys. Rev. Lett. **113**, 106802 (2014).

[42] J. P. Eisenstein, L. N. Pfeiffer, and K. W. West, Phys. Rev. Lett. **69**, 3804 (1992).

[43] H. Asahina and A. Morita, J. Phys. C: Solid State Phys. **17**, 1839 (1984).

[44] M. Bello, E. Levin, B. Shklovskii, and A. Efros, Sov. Phys. JETP **80**, 822 (1981).

[45] L. Wang *et al.*, Phys. Rev. B **89**, 075410 (2014).

# Supporting Information

## for

## Negative Compressibility in Graphene-terminated Black Phosphorus Heterostructures

Yingying Wu[1,†], Xiaolong Chen[1,2,†], Zefei Wu[1], Shuigang Xu[1], Tianyi Han[1], Jiangxiazi Lin[1], Yuan Cai[1], Yuheng He[1], Chun Cheng[3], Ning Wang*,[1]

[1]*Department of Physics and the William Mong Institute of Nano Science and Technology, the Hong Kong University of Science and Technology, Hong Kong, China*

[2] *Department of Engineering and Cambridge Graphene Centre, University of Cambridge, 9, JJ Thomson Avenue, Cambridge, United Kingdom, CB3 0FA*

[3] *Department of Materials Science and Engineering and Shenzhen Key Laboratory of Nanoimprint Technology, South University of Science and Technology, Shenzhen 518055, China*

*Correspondence should be addressed to: Ning Wang, phwang@ust.hk

**I. Fabrication of BN-BP-BN heterostructures**

To achieve ultraclean interfaces and highly stable black phosphorus (BP) devices in atmospheric conditions, we fabricate BN/BP/BN heterostructures and conduct high-temperature annealing to stabilize BP device structures. Polymer-free van der Waals transfer techniques [S[1]] are adopted as shown in Figure S1. First, thin flakes of graphene and BP are mechanically exfoliated onto a p-type silicon substrate covered with 300 nm thick $SiO_2$ separately. Then, the few-layer BP mechanically exfoliated onto another $SiO_2$/Si substrate is picked up by a thin BN (~4 nm-10 nm) flake. The BN-BP sample is used to pick up the graphene electrode (few-layer graphene). Finally, the BN-BP-Graphene structure is transferred onto a BN sheet supported on a $SiO_2$/Si substrate. The whole device structure is then dipped in Acetone to remove PDMS. In order to further improve the BP sample quality, as-fabricated devices are annealed at about 500°C in an argon (mixed with a few percent of hydrogen) atmosphere. The fabrication key step is to make an appropriate overlap (comparatively small) between graphene and BP. There should be an overlap between the top electrode and Graphene-BP contact area.

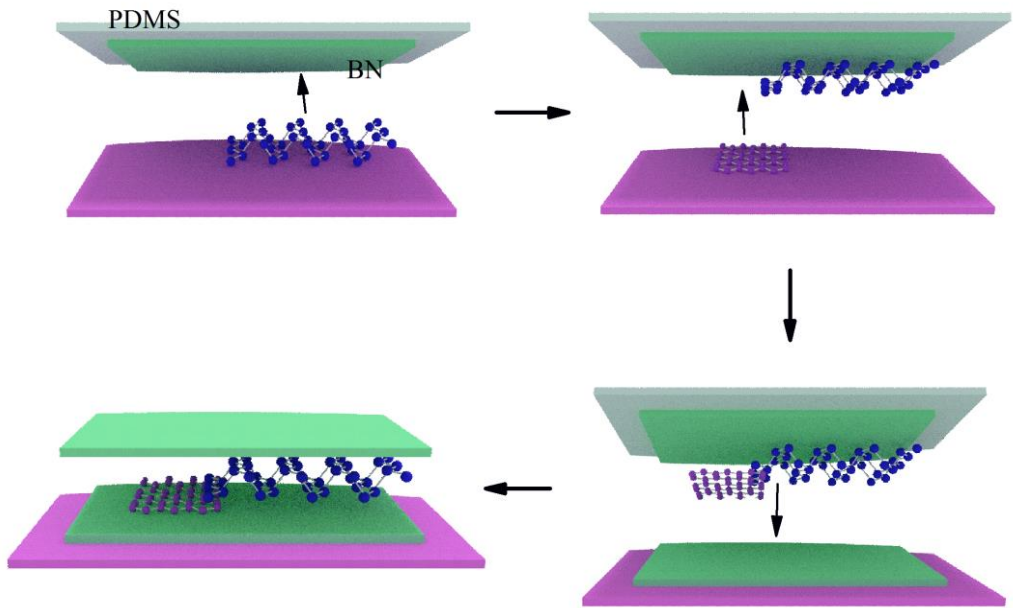

**Supplementary Fig. 1** Process flow of BN-BP-BN heterostructure fabrication.

## II. Charge traps in BP

High temperature annealing can significantly reduce the charge trap density in BP. As shown in Supplementary Figure 2a, there is no hysteresis effect. The data are collected from device C (13 nm thick BP) using an excitation voltage of 50mV.

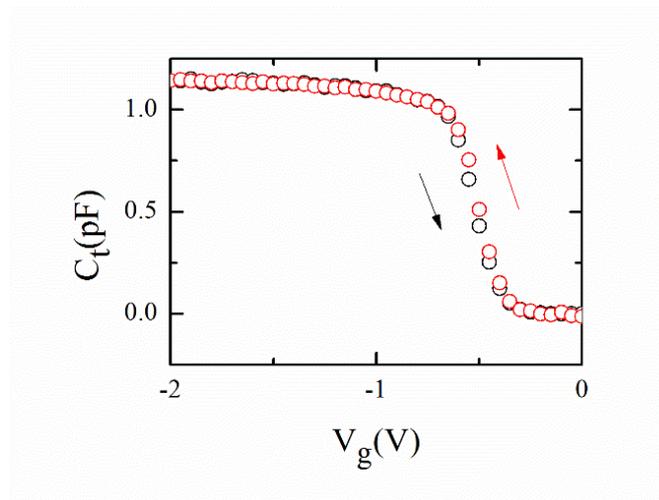

**Supplementary Fig. 2** No hysteresis effect is observed in 13 nm BP sample.

**III. Comparison of capacitor device structures**

To further illustrate the benefits of our device structure, we compare the performances of three different capacitor structures (one capacitor device with graphene electrode and two capacitors without using graphene terminal are shown in Supplementary Figure 3). To fabricate the device shown in Supplementary Figure 3b, a hard mask is defined by standard electron-beam lithography (using ZEP-520 resist). $O_2$-plasma is used to etch the top BN. Since the etching rates of $O_2$-plasma for BN and BP are different, BN layers can be quickly etched away while the BP layer still survives. Cr/Au electrodes are then deposited at the locations defined by electron-beam lithography. These steps guarantee polymer-free interfaces in BN-BP structures. However, the drawback is that a large resistance exists between two electrodes. The third method can eliminate the large resistance as shown in Supplementary Figure 3c since Cr/Au electrodes are deposited directly on BP before covering the top BN. However, the BN-BP interface is contaminated by polymers.

Using graphene as a terminal, we can simplify the fabrication and eliminate the chemical etching process. The interfaces in the BN-BP-BN structures are polymer-free as shown in Supplementary Figure 3a. By uncovering part of the few-layer graphene terminal, Cr/Au electrodes can be directly deposited on graphene without using the etching process. A small overlap between the top electrode and BP-Graphene contact area is necessary.

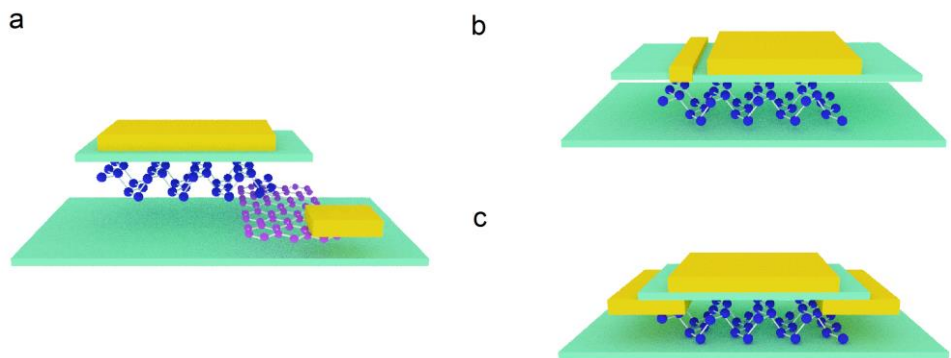

**Supplementary Fig. 3** Comparison between three capacitor configurations. (a) Structure used for the present devices. (b,c) Other two types of capacitor devices without using graphene terminals.

## IV. Temperature effects

Supplementary Figure 4 shows the capacitance of a 13 nm thick BP device. At 2K, the capacitance enhancement occurs when the excitation frequency is lower than 40Hz. The enhancement of capacitance at $nd^2 \to 0$, however, eventually disappears when temperature is increased to 20K.

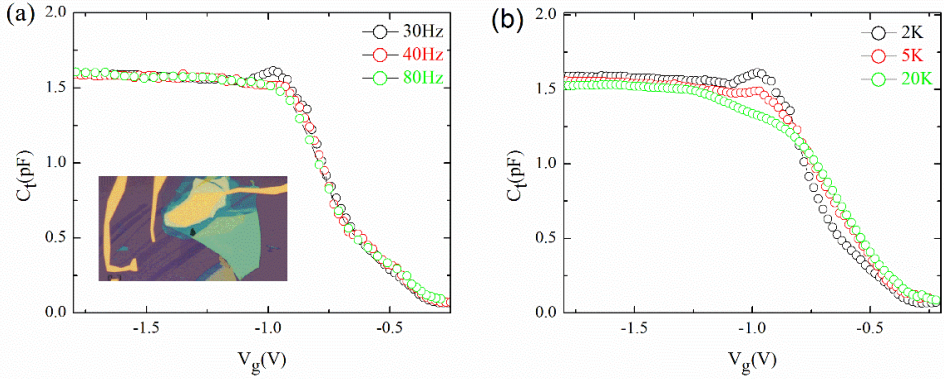

**Supplementary Fig. 4** Total capacitance measured from a 13 nm thick BP device. (a) Total capacitance measured at 2K with different excitation frequencies. (b) Total capacitance measured at different temperatures using a fixed excitation frequency of 30 Hz.

## V. Extraction of quantum capacitance

The measured capacitance can be described by $C_t' = C_t + C_r$, where $C_t$ is the capacitance shown in the main text $\frac{1}{C_t} = \frac{1}{C_g} + \frac{1}{C_q}$ and $C_r$ originates from the measurement setup. The parasitic capacitance $C_r$ can be accurately determined by a simple method[S2]. In the device structure without any overlap between the top electrode and bottom gate, there is a non-conductive channel in BP. This channel cannot be tuned by the bottom gate. The measured value of $C_r$ is in the order of ~$f$F, which is about 3-order smaller than the total capacitance. Subtracting the geometrical capacitance of BN (dielectric constant ~ 3.1), we extract the quantum capacitance.

**Reference**

[S1] Wang, L., et al. One-dimensional electrical contact to a two-dimensional material. *Science* 342.6158 (2013): 614-617.

[S2] Chen X, Wu Z, Xu S, et al. Probing the electron states and metal-insulator transition mechanisms in molybdenum disulphide vertical heterostructures[J]. *Nature communications* 6, (2015).